# Direct high-precision measurement of the magnetic moment of the proton

A. Mooser[1,2,a], S. Ulmer[3], K. Blaum[4], K. Franke[3,4], H. Kracke[1,2], C. Leiteritz[1], W. Quint[5,6],

C. C. Rodegheri[1,4], C. Smorra[3], J. Walz[1,2]

[1]Institut für Physik, Johannes Gutenberg-Universität Mainz, 55099 Mainz, Germany

[2]Helmholtz-Institut Mainz, 55099 Mainz, Germany

[3]RIKEN, Ulmer Initiative Research Unit, Wako, Saitama 351-0198, Japan

[4]Max-Planck-Institut für Kernphysik, 69117 Heidelberg, Germany

[5]Ruprecht-Karls-Universität Heidelberg, 69047 Heidelberg, Germany

[6]GSI - Helmholtzzentrum für Schwerionenforschung, 64291 Darmstadt, Germany

[a]present address: RIKEN, Ulmer Initiative Research Unit, Wako, Saitama 351-0198, Japan.

**The spin-magnetic moment of the proton $\mu_p$ is a fundamental property of this particle. So far $\mu_p$ has only been measured indirectly, analysing the spectrum of an atomic hydrogen maser in a magnetic field [1]. Here, we report the direct high-precision measurement of the magnetic moment of a single proton using the double Penning-trap technique [2]. We drive proton-spin quantum jumps by a magnetic radio-frequency field in a Penning trap with a homogeneous magnetic field. The induced spin-transitions are detected in a second trap with a strong superimposed magnetic inhomogeneity [3]. This enables the measurement of the spin-flip probability as a function of the drive frequency. In each measurement the proton's cyclotron frequency is used to determine the magnetic field of the trap. From the normalized resonance curve, we extract the particle's magnetic moment in units of the nuclear magneton $\mu_p = 2.792\,847\,350\,(9)\,\mu_N$. This measurement outperforms previous Penning trap measurements [4,5] in terms of precision by a factor of about 760. It improves the precision of the forty year old indirect measurement, in which significant theoretical bound state corrections [6] were required to obtain $\mu_p$, by a factor of 3. By application of this method to the antiproton magnetic moment $\mu_{\bar{p}}$, the fractional precision of the recently reported value [7] can be improved by a factor of at least 1000. Combined with the present result, this will provide a stringent test of matter/antimatter symmetry with baryons [8].**

The challenge to measure the properties of the proton with great precision inspires very different branches of physics. As part of the intense search for baryon number violation, to date, a lower limit of the particle's lifetime $t_p > 2.1 \cdot 10^{29}$ years is given [9]. Employing Penning traps, the proton's atomic mass was measured with a fractional precision of 0.14 parts in a billion (ppb) [10], and the proton-to-electron mass ratio was determined with a relative accuracy of 94 parts in $10^{12}$ [11]. Both provide essential input parameters to test the theory of quantum electrodynamics and contribute to the search for physics beyond the Standard Model. Furthermore, exciting results obtained by spectroscopy of muonic hydrogen [12] yielded a new value of the proton charge radius, and compared to previous measurements a 7 sigma deviation was observed, which has yet to be understood.

Another property of the proton is its spin magnetic moment

$$\mu_p = g_p \frac{q_p}{2m_p} S, \qquad (1)$$

where $q_p/m_p$ is the charge-to-mass ratio and $S$ is the particle's spin. The constant $g_p$ is a dimensionless measure of $\mu_p$ in units of the nuclear magneton $\mu_N = q_p \hbar/(2m_p)$. The most precise value of $\mu_p$ (see Fig. 1, white bar) is extracted from spectroscopy of atomic hydrogen [1]. In this experiment the bound proton-to-electron magnetic moment ratio $\mu_p(H)/\mu_e(H)$ was measured with a fractional precision of 10 ppb, and $\mu_p$ was calculated by taking theoretical corrections at the level of about 18 ppm into account [6].

A scheme for the direct measurement of magnetic moments of single particles in Penning traps has been applied with great success in measurements of the $g$-2 values of the electron and the positron [13] and further improved in [14], where fractional precisions at the level of 3.8 and 0.28 parts in $10^{12}$ were achieved, respectively. The application of this scheme to measure the magnetic moment of the proton is a considerable challenge, since $\mu_p$ is about 658 times smaller than that of the electron $\mu_e$. Thus, an apparatus with much higher sensitivity to the magnetic moment is needed.

In a Penning trap, the $g$-factor is determined by the measurement of a frequency ratio $g/2 = \nu_L/\nu_c$ where $\nu_c = q_p B_0/(2\pi m_p)$ is the cyclotron frequency, and $\nu_L = (g/2)\nu_c$ is the spin-precession frequency, also called the Larmor frequency. Both frequencies are measured in the same magnetic field $B_0$. The cyclotron frequency is obtained by the Brown-Gabrielse invariance theorem, $\nu_c^2 = \nu_+^2 + \nu_z^2 + \nu_-^2$, where $\nu_+$, $\nu_z$ and $\nu_-$ are the characteristic oscillation frequencies of the trapped particle [15], the modified cyclotron frequency, the axial frequency and the magnetron frequency, respectively. The Larmor frequency can be measured by application of the so-called continuous Stern-Gerlach-effect [3]. In that scheme a magnetic field inhomogeneity $\Delta B(z)_z = B_2 \cdot z^2$ is superimposed to the trap. This "magnetic bottle" couples the spin magnetic moment to the axial oscillation frequency, thus reducing the determination of the spin state to a measurement of $v_z$. A spin-flip shifts the axial frequency by $\Delta v_{z,SF} = (\mu_p B_2)/(2\pi^2 m_p v_z)$. This enables the measurement of the spin-transition rate as a function of an applied drive-frequency $v_{rf}$, and yields the Larmor frequency $v_L$ [16].

In our experiment, we use $B_2 = 2.97(10) \cdot 10^5$ T/m² [4], which is 2000 times stronger than in the electron/positron experiments [13]. In the presence of such a strong magnetic inhomogeneity the axial frequency shift caused by a spin-transition is $\Delta v_{z,SF} = 171$ mHz out of $v_{z,AT} \approx 740$ kHz. Thus, the detection of proton spin-quantum jumps requires an adequate axial frequency stability which is difficult to achieve in the strong magnetic bottle $B_2$ [17]. However, dramatic progress in the manipulation of a single trapped proton allowed the first direct measurements of the proton magnetic moment $\mu_p$ [4,5] (see Fig. 1, grey bars). Those experiments were carried-out solely in Penning traps with a superimposed $B_2$, which are usually called "analysis traps". The strong inhomogeneity broadens the width of the spin resonance, and ultimately limits the precision to the ppm level. An elegant solution to boost experimental precision is provided by the double Penning-trap technique [2]. This method separates the analysis of the spin state from the precision measurements of $v_c$ and $v_L$. In addition to the analysis trap (AT), a precision trap (PT) is used, in which the magnetic field is by orders of magnitude more homogeneous. This narrows the width of the Larmor resonance dramatically, and thus improves the precision. Here we report the first direct measurement of the proton magnetic moment using this technique.

Figure 2a.) shows a schematic of our double Penning-trap setup located at the University of Mainz, Germany. It is mounted in the horizontal and southward oriented bore of a superconducting magnet, with a magnetic field of $B_0 \approx 1.89$ T and a stability of $(\Delta B/B_0)/\Delta t = 4.0(0.7) \cdot 10^{-9}$/h. Each trap consists of five stacked cylindrical electrodes. The central ring electrode of the analysis trap is made out of ferromagnetic *Co/Fe* material, which generates the magnetic bottle. The other electrodes are made out of oxygen free copper. To prevent oxidation all electrodes are gold plated. The two trap centres are separated by 43.7 mm. Within this distance the magnetic field inhomogeneity drops significantly. In the precision trap we have $B_{PT} = B_0 + B_{1,PT} \cdot z + B_{2,PT} \cdot z^2$, where $B_{1,PT} \approx 85$ T/m and $B_{2,PT} \approx 4$ T/m$^2$, which is 75,000 times smaller than in the analysis trap. To shuttle the particle from one trap to the other, potential ramps are applied to transport electrodes located between the two traps. Radio-frequency (rf) drives applied to coils mounted close to each trap generate oscillating magnetic fields to drive proton spin-transitions. The entire setup is placed in a hermetically sealed vacuum chamber cooled to 4 K. In this chamber pressures below $10^{-14}$ Pa are achieved [18], providing single proton storage-times of at least one year.

Protons are produced with an in-trap electron impact ion source. Electrons from a field emission gun hit a polyethylene target. Sputtered atoms are ionized in the centre of the precision trap. From the loaded ion-cloud a single proton is prepared by using well-established techniques [19].

Once a single proton is prepared, the particle's modified cyclotron mode is cooled resistively. This is achieved by tuning a cryogenic tank circuit, which acts on resonance as a resistor, to the cyclotron mode at $\nu_+ = 28.9$ MHz [20]. Subsequently, the particle is shuttled to the analysis trap and the axial frequency is measured. To this end, $\nu_z \propto \sqrt{V_0}$ is tuned to resonance with our superconducting axial detection system at 740 kHz by adjusting the trapping voltage $V_0$. Once the axial motion is cooled to thermal equilibrium, the particle shorts the thermal noise of the detector and appears as a dip in the fast Fourier transform (FFT) of the detector signal [21]. Such an FFT-spectrum, recorded in 60s, is shown in Fig. 2b.), red data points. Applying a fit to the data the axial frequency $\nu_z$ is obtained. From this measurement, the quantum number of the cyclotron mode $n_+$ is determined [22]. Low $n_+$ is crucial to achieve an axial frequency stability which is sufficient to resolve single spin-transitions [23,24]. Below a threshold cyclotron quantum number $n_{+,th} < 1500$ we achieve single spin-flip resolution with a fidelity $F > 75\%$. This means that three out of four spin-transitions are identified correctly. If $n_+ > n_{+,th}$ is obtained, the particle is shuttled back to the precision trap and the cyclotron mode is cooled again. This procedure is repeated until $n_+ < n_{+,th}$, which takes about two hours.

Once a particle with adequate $n_+$ is prepared, the actual *g*-factor measurement is conducted. First, the proton's spin state is identified. To this end, the axial frequency is measured and a spin-transition is induced by applying a magnetic rf-drive, followed by another measurement of $v_z$. As soon as a spin quantum jump is observed (see Fig. 2c.)), the proton's spin state is known. Afterwards, the particle is transported to the precision trap where its cyclotron frequency $v_c$ is determined. First, the modified cyclotron frequency is measured via sideband coupling. An electric field at $v_{sb}$ which is close to $v_+ - v_{z,PT}$ is applied. This transfers energy between the axial mode and the modified cyclotron mode, and leads to a modulation of the axial oscillation amplitude. In the corresponding FFT spectrum a so-called "double-dip" with frequencies $v_l$ and $v_r$ is observed, as shown in Fig. 2b.), black data points. Subsequently, the axial frequency is recorded, which is about $v_{z,PT} \approx 624$ kHz. We determine $v_+$ by applying the relation [25]

$$v_+ = v_{sb} + v_l + v_r - v_{z,PT}. \tag{2}$$

The magnetron frequency $v_- \approx 7$ kHz is measured in a similar way. Finally, $v_{c,1}$ is obtained using the invariance theorem [15]. Next, spin-transitions are induced by applying a spin-flip drive at $v_{rf,PT}$ and subsequently the cyclotron frequency $v_{c,2}$ is measured again. Since the sideband drive leads to heating of the cyclotron mode, in a next step, by repeated cyclotron mode cooling in the precision trap and transport to the analysis trap, a low $n_+ < n_{+,th}$ state is prepared and the spin state analysed again. From a comparison of the two measured spin states we conclude whether the spin in the precision trap was flipped. By repeating this scheme many times the spin-flip probability $P_{SF}$ as a function of $v_{rf,PT}$ is obtained. Normalizing each applied $v_{rf,PT,k}$ by the associated $v_{c,k}$ where $k$ is the measurement number, one obtains a so-called $g$-factor resonance, $P_{SF}(v_L/v_c)$, with a maximum at $\mu_p/\mu_N = g_p/2$. For the normalization we use the average $(v_{c,1}+v_{c,2})/2$ of the two cyclotron frequency measurements. This compensates for linear magnetic field drifts which potentially take place while spin-transitions are driven.

The result of our $g$-factor measurement is shown in Fig. 3. Incoherences caused by the coupling of the particle's axial motion to the axial detection circuit in the slightly inhomogeneous magnetic field of the precision trap prevent $P_{SF}$ from exceeding 50% [16]. The line-width of the measured resonance is 12.5 ppb, which is caused by saturation and thermal fluctuations of the modified cyclotron energy $E_+$. The latter causes fluctuations of the axial frequency, $\nu_{z,PT} \propto B_2 \cdot E_+$, which lead to fluctuations of the measured cyclotron frequency via Eq. 2, thus contributing to the line width. The measured dataset is analysed using the maximum likelihood method [26] based on a Gaussian distribution with the $g$-factor being the line shape centre. The maximum likelihood method is a statistical parameter estimation technique and avoids the need for arbitrary data binning, which may result in a biased estimate of the fitting parameters. From the data analysis we obtain $\mu_p = \frac{g_p}{2}\mu_N = 2.792\,847\,348\,(7)\mu_N$, where the number in parenthesis is the statistical uncertainty of the fit, which corresponds to a relative precision of 2.6 ppb.

Systematic shifts $\Delta(g_p/2)$ of the measured $(g_p/2)$-value are caused by time and energy dependencies of $\nu_L$ and $\nu_c$

$$\frac{\Delta(g_p/2)}{(g_p/2)} = \frac{\Delta\nu_L(E_+,E_z,E_-,t)}{\nu_L} - \frac{\Delta\nu_C(E_+,E_z,E_-,t)}{\nu_C} \quad (3)$$

The frequency shifts $\Delta\nu_L(E_+,E_z,E_-,t)$ and $\Delta\nu_c(E_+,E_z,E_-,t)$ are caused by trap imperfections such as a slightly inhomogeneous magnetic field at the centre of the precision trap or an anharmonic trapping potential. To first order, the shifts induced by the magnetic field inhomogeneities cancel in the frequency ratio, since they contribute the same relative amount to $\nu_c$ and $\nu_L$ [4,15]. A considerable systematic shift can be caused by an anharmonicity of the electrostatic potential which only affects $\nu_c$ while $\nu_L$ remains unchanged. Thus, the trapping potential was optimized by careful adjustment of the voltages applied to the compensation electrodes of the precision trap to obtain $\Delta\nu_c = 0$. The relative uncertainty in the resulting shift $\Delta\nu_c$ is 0.20 ppb. A second $g$-factor resonance was recorded where the electrostatic potential was deliberately de-tuned to shift the modified cyclotron frequency by −5 ppb. Within error-bars we obtained the same $(g_p/2)$-value, which confirms that systematic shifts due to electrostatic anharmonicities are understood and negligible at the present level of precision. In addition to these two leading-order shifts, relativistic frequency shifts and image-charge shifts contribute [15]. However, since the frequency measurements are carried out in thermal equilibrium with the cryogenic detection system the relativistic shift contributes only at a level of $\left(\Delta(g_p/2)\right)_{rel} = 0.03$ ppb. For our trap geometry the image-charge shift is $\left(\Delta(g_p/2)\right)_{ic} = -0.088$ ppb. Thus, both shifts can be neglected. Time dependent shifts are due to voltage and magnetic field drifts. The effect of voltage drifts was characterized by performing a sequence of axial frequency measurements at constant $E_+$. The corresponding systematic shift in $v_c$ is –0.07(0.35) ppb.

The dominant systematic uncertainty is caused by nonlinear drifts of the magnetic field. By comparing the cyclotron frequency measurements before and after application of the spin-flip drive we find an average shift $v_{c,2}$-$v_{c,1}$ = 4 ppb. Such frequency shifts in the precision trap are observed only if the spin-flip drive in the analysis trap has been applied in advance. It is consistent with heating of the electrodes caused by the latter spin-flip drive. Due to thermal expansion the trap centres are shifted, thereby changing the magnetic field in the precision trap. Once the drive is turned off, thermal relaxation causes the observed drift of the magnetic field. The last contribution considered is a shift of the axial frequency after the measurement of $v_+$. The sideband drive heats the particle to a cyclotron energy of $E_+/k_B = T_+ \approx 330$ K. During the subsequent axial frequency measurement, the modified cyclotron mode is cooled resistively. This leads to an effective frequency shift of $v_z$ and contributes to a shift of the cyclotron frequency by –0.51(0.08) ppb. Accordingly, the magnetic moment value is corrected and the final result is

$$\frac{\mu_p}{\mu_N} = \frac{g_p}{2} = 2.792\ 847\ 350\ (7)(6). \qquad (4)$$

The first and second numbers in parenthesis are the statistical uncertainty of the fit and the systematic uncertainty, respectively, see Table 1. The latter is dominated by the nonlinear drift of the magnetic field. The result has a relative precision of 3.3 ppb and is in agreement with the currently accepted CODATA-value $g_{CODATA}/2 = 2.792\ 847\ 356(23)$, but is 2.5 times more precise.

We expect that at least an improvement by another factor of 10 in precision will be possible, with an apparatus with reduced residual magnetic field inhomogeneity $B_2$ in the precision trap, higher spin state detection fidelity as well as application of phase-sensitive detection techniques [27]. In addition, the so-called Standard Model extension [8] describes diurnal frequency variations as a consequence of CPT and Lorentz violation. With faster measurement cycles, which become possible by application of advanced cyclotron cooling techniques, and an improved spin state detection fidelity a search for the predicted effects [28] becomes feasible.

The double Penning-trap method can be applied to measure the antiproton magnetic moment with similar precision [7,29]. A comparison of both values will provide a sensitive test of CPT-invariance with baryons. The measurement of the antiproton magnetic moment will be conducted in the BASE experiment [30] at the Antiproton Decelerator of CERN.

*References:*

[1] Winkler, P. F., Kleppner, D., Myint, T. & Walther, F. G. Magnetic moment of the proton in Bohr magnetons. Phys. Rev. A **5**, 83-114 (1972).

[2] Häffner, H. *et al*. Double Penning-trap technique for precise *g*-factor determinations in highly charged ions. Eur. Phys. J. D **22**, 163-182 (2003).

[3] Dehmelt, H. G. & Ekström, P. Proposed *g*-2 experiment on single stored electron or positron. P. Bull. Am. Phys. Soc. **18**, 727-731 (1973).

[4] Rodegheri, C. C. *et al*. An experiment for the direct determination of the *g*-factor of a single proton in a Penning trap. New J. Phys. **14**, 063011 (2012).

[5] DiSciacca, J. & Gabrielse, G. Direct measurement of the proton magnetic moment. Phys. Rev. Lett. **108**, 153001 (2012).

[6] Karshenboim, S. G. & Ivanov, V. G. The *g*-factor of the proton. Phys. Lett. B **566**, 27-34 (2003).

[7] DiSciacca, J. *et al*. One-particle measurement of the antiproton magnetic moment. Phys. Rev. Lett. **110**, 130801 (2013).

[8] Bluhm, R., Kostelecky, V. A. & Russell, N. CPT and Lorentz tests in Penning traps. Phys. Rev. D **57**, 3932-3943 (1998).

[9] Beringer, J. *et al.* (Particle Data Group). Review of particle physics. Phys. Rev. D **86**, 010001 (2012).

[10] Van Dyke, R. S., Farnham, D. L., Zafonte, S. L. & Schwinberg, P. B. High precision Penning trap mass spectroscopy and a new measurement of the proton's "atomic mass". AIP Conf. Proc. **457**, 101-110 (1999).

[11] Sturm, S. *et al.* High-precision measurement of the atomic mass of the electron. Nature **506**, 467-470 (2014).

[12] Pohl, R. *et al*. The size of the proton. Nature **466**, 213-216 (2010).

[13] Van Dyck, R. S., Schwinberg, P. B. & Dehmelt, H. G. New high-precision comparison of electron and positron *g*-factors. Phys. Rev. Lett. **59**, 26-29 (1987).

[14] Hanneke, D., Fogwell, S. & Gabrielse, G. New measurement of the electron magnetic moment and the fine structure constant. Phys. Rev. Lett. **100**, 120801 (2008).

[15] Brown, L. S. & Gabrielse, G. Geonium theory: physics of a single electron or ion in a Penning trap. Rev. Mod. Phys. **58**, 233-311 (1986).

[16] Brown, L. S. Geonium Lineshape. Ann. Phys. **159**, 62-98 (1985).

[17] Ulmer, S. *et al*. Observation of spin flips with a single trapped proton. Phys. Rev. Lett. **106**, 253001 (2011).

[18] Gabrielse, G. *et al*. Special relativity and the single antiproton: fortyfold improved comparison of $\bar{p}$ and p charge-to-mass ratios. Phys. Rev. Lett **74**, 3544-3547 (1995).

[19] Ulmer, S. *et al*. Direct measurement of the free cyclotron frequency of a single particle in a Penning trap. Phys. Rev. Lett. **107**, 103002 (2011).

[20] Ulmer, S. *et al.* A cryogenic detection system at 28.9 MHz for the non-destructive observation of a single proton at low particle energy. Nucl. Instrum. Meth. A **705**, 55-60 (2013).

[21] Wineland, D. J. & Dehmelt, H. G. Principles of the stored ion calorimeter. J. Appl. Phys. **46**, 919-930 (1975).

[22] Mooser, A. *et al.* Demonstration of the double Penning-trap technique with a single proton. Phys. Lett. B **723**, 78-81 (2013).

[23] Mooser, A. *et al.* Resolution of single spin-flips of a single proton. Phys. Rev. Lett. **110**, 140405 (2013).

[24] DiSciacca, J., Marshall, M., Marable, K. & Gabrielse G. Resolving an individual one-proton spin flip to determine a proton spin state. Phys. Rev. Lett. **110**, 140406 (2013).

[25] Cornell, E. A., Weisskoff, R. M., Boyce, K. R & Pritchard, D. E. Mode coupling in a Penning trap: π pulses and a classical avoided crossing. Phys. Rev. A **41**, 312-315 (1990).

[26] Sivia, D. S. & Skilling J. Data Analysis - A Bayesian tutorial, Oxf. Sci. Pub., Edt. 1 (2010).

[27] Sturm, S. *et al.* g-factor measurement of hydrogenlike $^{28}$Si$^{13+}$ as a challenge to QED calculations. Phys. Rev. A **87**, 030501 (2013).

[28] Mittleman, R. K., Ioannou, I. I., Dehmelt, H. G. & Russell, N. Bound on CPT and Lorentz symmetry with a trapped electron. Phys. Rev. Lett. **83**, 2116-2119 (1999).

[29] Smorra, C. *et al.* Towards a high-precision measurement of the antiproton magnetic moment. Hyperfine Interact. DOI: 10.1007/s10751-014-1018-7 (2014).

[30] Ulmer, S. *et al.* Technical Design Report BASE, CERN Document Server, SPSC-TDR-002 (2013).

**Table 1| Error budget of the direct proton *g*-factor measurement.** The table lists the relative systematic shifts and their uncertainties, which were applied to the measured *g*-value.

**Figure 1| Relative precision achieved in measurements of the proton magnetic moment**. The value extracted indirectly from measurements with a hydrogen maser has a precision of 10 ppb [1]. Direct measurements with a single proton in a Penning trap with strong superimposed magnetic inhomogeneities were performed in 2012 by us [4] and a group at Harvard [5]. The result of the measurement described in this work was achieved by using the double Penning-trap technique with a single trapped proton. Our result is a factor of 3 times more precise than reported in [1] and about 760 times more precise than other direct single particle measurements.

**Figure 2| Experimental setup and measurement procedures**. a.) Schematic of the double Penning-trap setup which is used for the direct measurement of the proton magnetic moment. It consists of two traps, an analysis trap and a precision trap, which are connected by transport electrodes. A strong magnetic field inhomogeneity is superimposed to the analysis trap, which is required to detect proton spin-quantum-transitions. In the precision trap, where the magnetic field is homogeneous, the precise frequency measurements are carried out. The solid curve in the plot below the trap system indicates the strength of the on-axis magnetic field. b.) Fast Fourier Transform spectrum of the axial detector's output signal. The dip (red) is due to a single particle, which shorts the thermal noise of the detector. The double-dip signal (black) appears when a quadrupolar sideband drive at $v_+ - v_z$ is applied. From such dip-spectra $v_+$, $v_z$ and $v_-$ are determined and thus the cyclotron frequency. The axial frequency has been offset by 623,850 Hz. For further details see text. c.) Axial frequency measurement as a function of time. Frequency jumps of about 170 mHz are observed which are due to induced single proton spin-transitions. The axial frequency has been offset by 742,060 Hz.

**Figure 3| Measured *g*-factor resonance.** The abscissa is the measured *g*-value normalized to the currently accepted value $g_{CODATA}$. The solid line is a maximum-likelihood fit to the data which avoids the need for data binning. The shaded area indicates the $1\sigma$ confidence band of the fit. Filled squares representing binned data points with $1\sigma$ error bars are shown for visualization and do not explicitly enter the line fit. It took about 4 months to record the entire set of 450 data points.

***Acknowledgements:*** We acknowledge financial support of the BMBF, and the EU (ERC Grant No. 290870-MEFUCO), the Helmholtz-Gemeinschaft, HGS-HIRE, the Max-Planck Society, IMPRS-PTFS, and the RIKEN Initiative Research Unit Program.

***Author Contributions:*** S.U., C.C.R, H.K., A.M., and C.L. designed and built the experimental apparatus and the data acquisition system. A.M., C.L. and S.U. took part in the months-long data-taking runs. K.F. and A.M. developed the algorithms for the spin state analysis. A.M., K.F., S.U., C.L. and H.K. analysed the data. S.U., A.M., K.B. and J.W. wrote the initial manuscript which was then read, improved and finally approved by all authors.

***Author Information***: Reprints and permissions information is available at www.nature.com/reprints. The authors declare no competing financial interests. Correspondence and requests for materials should be addressed to A.M., mooser@uni-mainz.de.

| Parameter | Relative Shift of $g_p/2$ | Uncertainty |
|---|---|---|
| Trapping Potential | 0 | $0.2 \cdot 10^{-9}$ |
| Relativistic Shift | $0.03 \cdot 10^{-9}$ | - |
| Image-Charge Shift | $-0.088 \cdot 10^{-9}$ | - |
| Voltage Fluctuations | $-0.07 \cdot 10^{-9}$ | $0.35 \cdot 10^{-9}$ |
| Magnetic Field Relaxation | 0 | $2 \cdot 10^{-9}$ |
| Cyclotron Cooling | $-0.51 \cdot 10^{-9}$ | $0.08 \cdot 10^{-9}$ |
| Total Systematic Shift | $-0.64 \cdot 10^{-9}$ | $2 \cdot 10^{-9}$ |

**Table 1| Error budget of the direct proton *g*-factor measurement.**

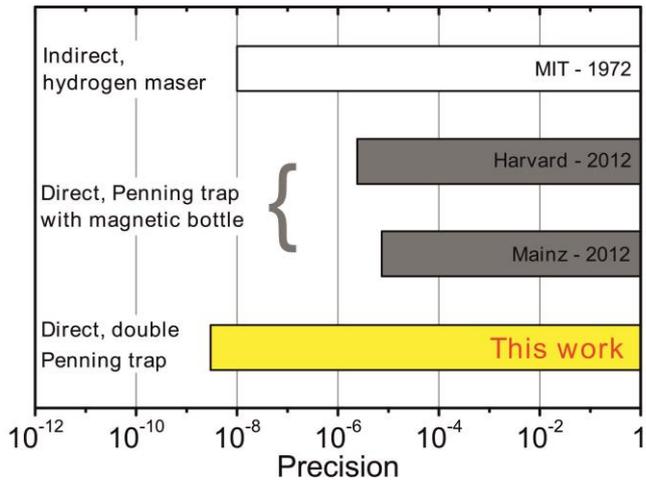

**Figure 1| Relative precision achieved in measurements of the proton magnetic moment**.

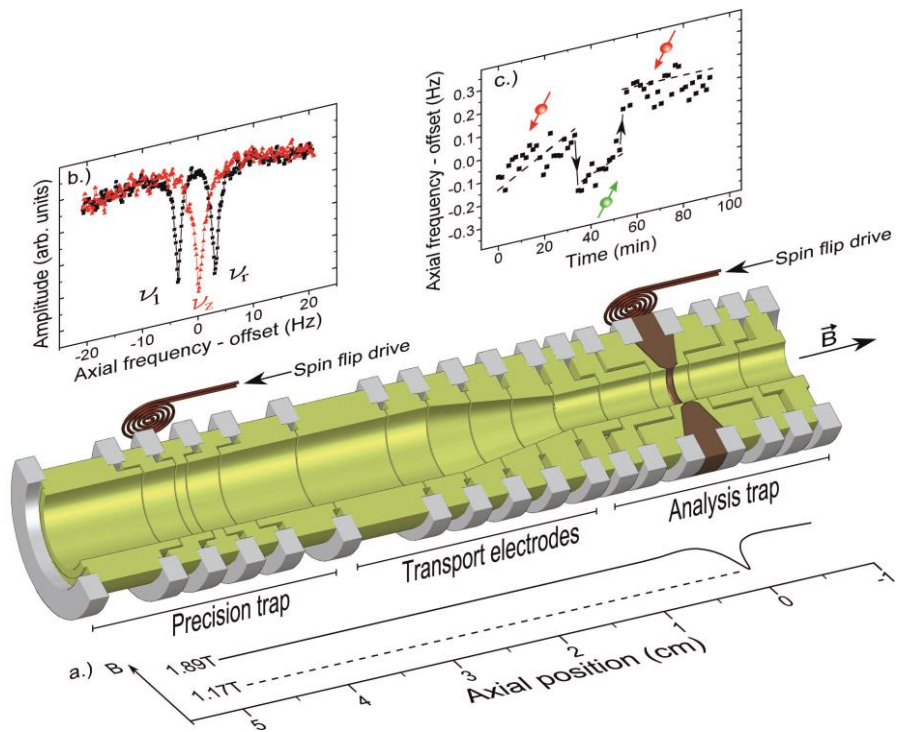

**Figure 2| Experimental setup and measurement procedures.**

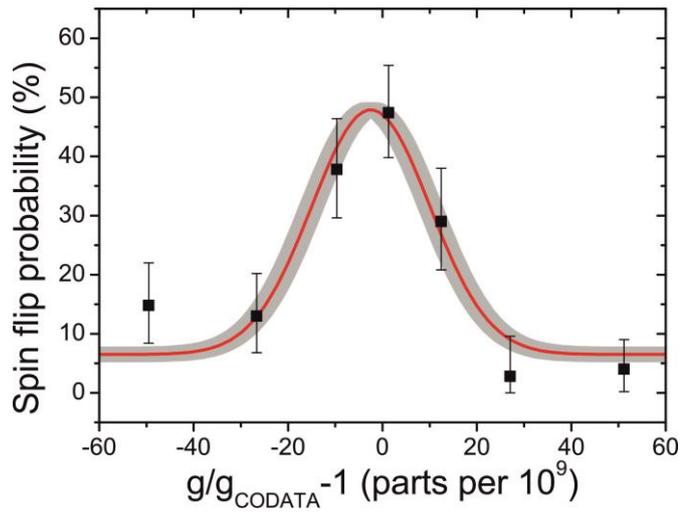

**Figure 3| Measured *g*-factor resonance.**